\documentclass[aps,prl,twocolumn,superscriptaddress,showpacs,floatfix]{revtex4}

\usepackage{graphicx}

\begin{document}


\title{Crossover and Coexistence of Quasiparticle Excitations in the Fractional Quantum Hall Regime at $\nu \leq 1/3$}


\author{C.F. Hirjibehedin}
\affiliation{Department of Physics, Columbia University, New York,
NY  10027} \affiliation{Bell Labs, Lucent Technologies, Murray
Hill, NJ 07974}

\author{A. Pinczuk}
\affiliation{Department of Physics, Columbia University, New York,
NY  10027} \affiliation{Bell Labs, Lucent Technologies, Murray
Hill, NJ  07974} \affiliation{Department of Appl. Physics and
Appl. Mathematics, Columbia University, New York, NY  10027}

\author{B.S. Dennis}
\affiliation{Bell Labs, Lucent Technologies, Murray Hill, NJ
07974}

\author{L.N. Pfeiffer}
\affiliation{Bell Labs, Lucent Technologies, Murray Hill, NJ
07974}

\author{K.W. West}
\affiliation{Bell Labs, Lucent Technologies, Murray Hill, NJ
07974}

\date{May 20, 2003}

\begin{abstract}

New low-lying excitations are observed by inelastic light scattering
at filling factors $\nu=p/(\phi p \pm 1)$ of the fractional quantum
Hall regime with $\phi=4$. Coexisting with these modes throughout the
range $\nu \leq 1/3$ are $\phi=2$ excitations seen at $1/3$. Both
$\phi=2$ and $\phi=4$ excitations have distinct behaviors with
temperature and filling factor. The abrupt first appearance of the
new modes in the low energy excitation spectrum at $\nu \lesssim 1/3$
suggests a marked change in the quantum ground state on crossing the
$\phi=2 \rightarrow \phi=4$ boundary at $\nu = 1/3$.

\end{abstract}


\maketitle


In the regime of the fractional quantum Hall (FQH) effect fundamental
interactions within a 2D electron system cause condensation into
quantum liquids. The fluids become incompressible at certain rational
values of the Landau level (LL) filling factor $\nu=n h c / e B$,
where $n$ is the areal density and B is the perpendicular magnetic
field. When only the lowest electron LL is occupied ($\nu \leq 1$),
the dominant FQH states are part of the sequences $\nu = p / (\phi p
\pm 1)$, where $p$ is an integer that enumerates members of a
particular sequence and $\phi$ is an even integer that labels
different sequences. The states at $\nu=1/(\phi \pm 1)$ are
boundaries that separate major sequences of the FQH effect. The
strongest such state occurs at $\nu=1/3$, which can be formed with
either $\phi=2$ or $\phi=4$. Magnetotransport
\cite{Jiang1990,Sajoto1990,Du1993,Leadley1994,Du1996,Pan2000} as well
as calculations
\cite{Girvin1985,Nakajima&Aoki,Kamilla&Jain,Scarola2000b} show that
excitations at FQH states with $\nu \geq 1/3$ ($\phi=2$) have
characteristic energy scales larger than those of states with $\nu <
1/3$ ($\phi=4)$.
\par

By associating strongly interacting electrons with an even number of
vortices $\phi$ of the many-body wavefunction, the composite fermion
($CF$) framework posits the existence of quasiparticles that
experience an effective magnetic field $B^* = \pm B / (\phi p \pm 1)$
that is reduced from $B$ due to Chern-Simons gauge fields that
account for Coulomb interactions
\cite{Kalmeyer&Zhang1992,Lopez&Fradkin1993,HalperinLeeRead}. Vortex
attachment and gauge fields result in weakly interacting $^{\phi}CFs$
(quasiparticles associated with $\phi$ vortices) that populate
quasi-LLs and exhibit an integer quantum Hall effect with effective
filling factor $\nu^*=p$ at exactly the FQH filling factors
$\nu=p/(\phi p \pm 1)$. The overlap of the FQH sequences at boundary
states $\nu=1/(\phi \pm 1)$ is present in the $CF$ hierarchy: the
$1/3$ state can be thought of as a $^2CF$ and $^4CF$ system, both at
$CF$ filling factor $\nu^*=1$ but each with opposite effective
magnetic field.
\par

Because $^2CF$ and $^4CF$ excitations have different energy scales,
new low-energy modes of $^4CF$ collective excitations could emerge at
$\nu \leq 1/3$. Resonant inelastic light scattering in the presence
of weak residual disorder offers direct access to the low energy
excitation spectrum of FQH liquids
\cite{Pinczuk1993,Davies1997,Kang2001}. In this Letter we report
light scattering experiments that probe quasiparticle crossover
between different $\phi$ sequences. Light scattering is here a
powerful tool because it accesses low energy excitations both at and
between the FQH states \cite{Dujovne2003}.
\par

We investigate the crossover between FQH sequences with a boundary at
$\nu=1/3$. Starting at filling factors slightly below $1/3$, we find
the emergence of new low lying excitations that reveal an interaction
energy scale significantly below that of the liquid at $\nu \gtrsim
1/3$. The existence of these modes throughout the $\phi=4$ regime as
well as their low energy and sharp temperature dependence suggests
they are excitations linked to $\phi=4$ quasipartices in the FQH
effect at $\nu \leq 1/3$.
\par

Remarkably, spin conserving quasiparticle-quasihole $^2CF$
excitations seen at $\nu \geq 1/3$ are also found for $\nu < 1/3$.
These $^2CF$ modes remain strong throughout the $\phi=4$ regime and
shift continuously in energy with magnetic field. Both $^2CF$ and
$^4CF$ excitations are found to coexist throughout the $\phi=4$
regime, as illustrated in the inset to Fig. \ref{fig:figure3}. These
modes have distinct behaviors with temperature and magnetic field,
confirming that they are related to excitations of different
quasiparticles. Excitations in the quantum Hall regime are
constructed from neutral pairs composed of quasiparticles excited to
a higher energy level and quasiholes that appear in the ground state
\cite{Kallin&Halperin1984,Haldane&Rezayi,Girvin1985}. The
simultaneous observation of $^2CF$ and $^4CF$ excitations is evidence
that multiple quasiparticle flavors can be created in the ground and
excited states.
\par

The $\phi=4$ modes are seen at filling factors that asymptotically
approach $1/3$ from below and remain with similar spectral intensity
throughout the range $1/3 \gtrsim \nu \gtrsim 2/7$. These low-lying
modes are first detected at the filling factor $\nu \lesssim 1/3$
that coincides with the disappearance of the long wavelength modes of
the $1/3$ state \cite{Pinczuk1993,Davies1997}. The appearance of the
new low energy modes and the coexistence with remaining $\phi=2$
modes at $\nu \lesssim 1/3$ mark a dramatic transition in quantum
ground state properties of the 2D electron system when crossing the
boundary at $1/3$.  The new ground state at $\nu < 1/3$ incorporates
both $\phi=2$ and $\phi=4$ quasiparticle excitations. Among the
intriguing possibilities for the new ground state is that of distinct
coexisting $\phi=2$ and $\phi=4$ liquids.
\par

\begin{figure}
\includegraphics{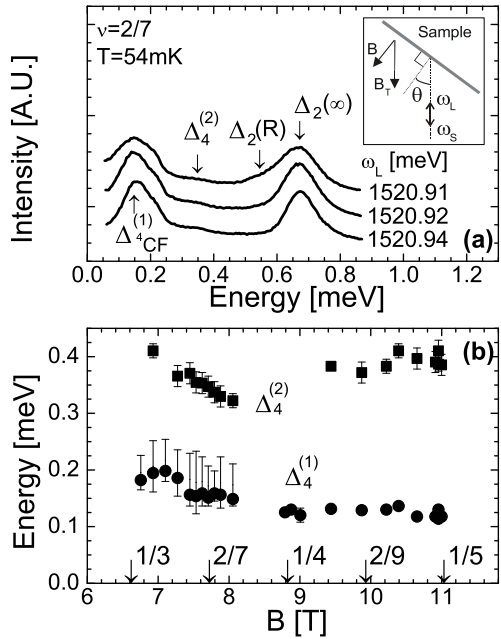}
\caption{\label{fig:figure1} (a) Light scattering spectra at $\nu =
2/7$ and $T = 54mK$ for several $\omega_L$. Various labelled
excitations, discussed in the text, are resonant at different
energies. The inset shows the backscattering geometry. (b) Magnetic
field dependence of the peak energy of $\Delta_{4}^{(1)}$ (solid
circles) and $\Delta_{4}^{(2)}$ (solid squares). Filling factors are
labelled at the bottom.}
\end{figure}

The 2D electron system studied here is formed in a GaAs single
quantum well with width $w = 330\AA$. The electron density is $n=5.4
\times 10^{10}cm^{-2}$ and the low temperature mobility is $\mu = 7.2
\times 10^6 cm^2/Vs$. Light scattering measurements at a base
temperature of 50mK  are performed in a dilution refrigerator through
windows for direct optical access. The energy of the incident photons
$\omega_L$ is tuned close to the fundamental optical gap of the GaAs
well and the power density is kept below $10^{-4} W/cm^2$. The sample
is mounted in a backscattering geometry, making an angle $\theta$
between the incident photons and the normal to the sample surface, as
illustrated in the inset of Fig. \ref{fig:figure1}a. The magnetic
field perpendicular to the sample is $B = B_T cos \theta$, where
$B_T$ is the total applied field. The wavevector transferred from the
photons to the 2D system is $k=(2 \omega_L / c) sin \theta$.
\par

\begin{figure}
\includegraphics{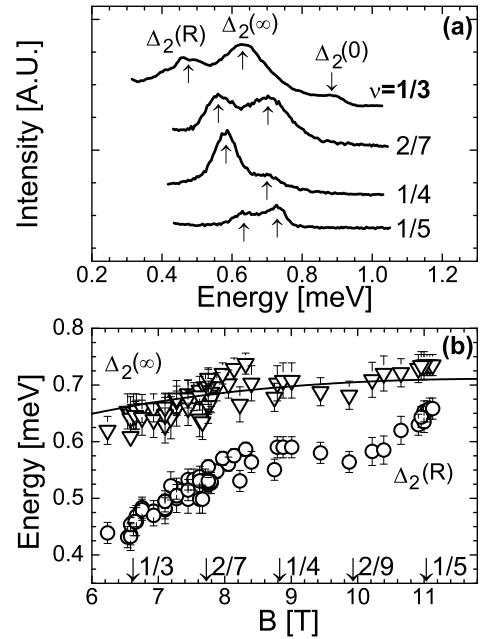}
\caption{\label{fig:figure2} (a) Spectra at $\nu < 1/3$ and $T \sim
50mK$. Up-arrows mark $\Delta_{2}(R)$ and $\Delta_{2}(\infty)$
excitations while the down-arrow indicates the $\Delta_{2}(0)$ mode
at $\nu=1/3$. (b) Magnetic field dependence of $\Delta_{2}(R)$ (open
circles) and $\Delta_{2}(\infty)$ (open triangles). The solid line is
a fit of $\Delta_{2}(\infty)$ that includes a finite width
correction, as described in the text. Filling factors are labelled at
the bottom.}
\end{figure}

Figure \ref{fig:figure1}a shows low-energy spectra at various
$\omega_L$ for $\nu = 2/7$. New excitations labelled
$\Delta_{4}^{(1)}$ and $\Delta_{4}^{(2)}$ are at energies below and
above the Zeeman energy $E_Z = g \mu_B B_T$, where $g=0.44$ is the
Lande g-factor for electrons in GaAs and $\mu_B$ is the Bohr
magneton. Two higher energy modes seen at $\nu \leq 1/3$ and labelled
$\Delta_{2}(R)$ and $\Delta_{2}(\infty)$ are discussed below. The
excitations $\Delta_{4}^{(1)}$ and $\Delta_{4}^{(2)}$ exist both at
and between the FQH states for $\nu < 1/3$ but are not seen at $\nu
\geq 1/3$. The mode $\Delta_{4}^{(1)}$ is strongest in the range $1/3
> \nu \gtrsim 2/7$, where its intensity does not qualitatively change
with $\nu$. Figure \ref{fig:figure1}b shows the field dependence of
the peak energies. Remarkably, the lower energy $\Delta_{4}^{(1)}$
modes do not have a strong dependence on magnetic field, shifting to
lower energy as the effective filling factor decreases between $1/3$
and $2/7$ and then remaining relatively constant for $1/5 \leq \nu
\lesssim 1/4$. The energies of $\Delta_{4}^{(2)}$ excitations have a
stronger dependence on magnetic field, decreasing as $\nu \rightarrow
1/4$ from both sides.
\par

Three modes at $\nu = 1/3$ shown in Fig. \ref{fig:figure2}a are
identified with spin-conserving $^2CF$ excitations. The mode labelled
$\Delta_{2}(0)$ is assigned to the wavevector conserving ($q = k
\lesssim 1 / 10 l_0 $) excitation and is seen only in a small range
of $\nu$ around $1/3$. At $\nu=1/3$, breakdown of wavevector
conservation activates excitations from the roton ($\Delta_{2}(R)$)
and large-q ($\Delta_{2}(\infty)$) critical points of the dispersion
of spin conserving excitations
\cite{Pinczuk1988,Davies1997,Kang2001}. These excitations have been
seen for a wider range in $\nu$ and were found to disappear between
$2/5 > \nu > 1/3$ \cite{Davies1997,Dujovne2003}. Surprisingly, as
shown in Figs. \ref{fig:figure1}a and \ref{fig:figure2}a, we find
that these two excitations exist throughout the $\phi=4$ regime of
$1/3
> \nu \geq 1/5$. Fig. \ref{fig:figure2}b shows the magnetic field
dependence of excitations $\Delta_{2}(R)$ and $\Delta_{2}(\infty)$
that exist throughout the range $1/3 \geq \nu \geq 1/5$. Because
these excitations evolve continuously from $\Delta_{2}(R)$ and
$\Delta_{2}(\infty)$ at $\nu = 1/3$, these modes are interpreted as
$^2CF$ excitations that continue to exist in the $\phi=4$ regime.
\par

\begin{figure}
\includegraphics{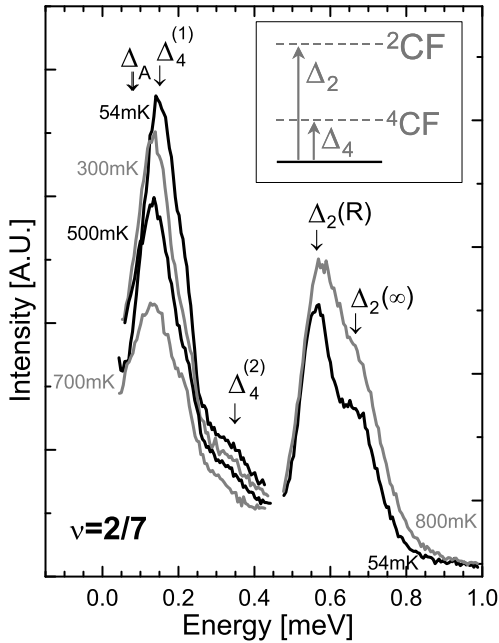}
\caption{\label{fig:figure3} Temperature dependence of $^2CF$
($\Delta_{2}(R)$ and $\Delta_{2}(\infty)$) and $^4CF$
($\Delta_{4}^{(1)}$ and $\Delta_{4}^{(2)}$) modes at $\nu=2/7$ and
the activation gap $\Delta_A$ \cite{Du1993,Pan2000}, scaled as
described in the text. The inset is a schematic of the energy level
structure corresponding to the transitions to which the modes are
assigned.}
\end{figure}

$\Delta_{4}^{(1)}$ and $\Delta_{4}^{(2)}$ excitations have the marked
temperature dependence shown in Fig. \ref{fig:figure3} at $\nu=2/7$.
We see that both modes begin to weaken above $T = 300mK$ and are
significantly lower in intensity at higher $T$. The modes associated
with $\Delta_{2}$ transitions, at $\Delta_{2}(R)$ and at
$\Delta_{2}(\infty)$ have markedly weaker temperature dependence.
Similar results are found throughout the $\phi=4$ regime.
\par

Because they exist only in the range $1/3 > \nu \geq 1/5$, it is
natural to associate the modes labelled $\Delta_{4}$ with excitations
of $\phi=4$ quasiparticles. Since the energy of the modes does not
increase with magnetic field, which would be characteristic of spin
excitations, these modes are identified as spin conserving
excitations due to the ${^4}CF$ transitions shown in the inset to
Fig. \ref{fig:figure3}. Activation gaps $\Delta_A$ for $\phi=4$ FQH
states at $2/7$ and $1/5$
\cite{Jiang1990,Sajoto1990,Du1993,Du1996,Pan2000}, scaled by the
Coulomb energy to match the density of our sample, are roughly
$0.08meV \sim 900mK$. As shown in Fig. \ref{fig:figure3}, $\Delta_A$
at $\nu=2/7$ is slightly lower than the peak energy of
$\Delta_{4}^{(1)}$ but is within the width of the peak and is
consistent with the temperature dependence. Similar differences are
found between the $\Delta_{2}(\infty)$ energy and the activation gap
at $1/3$ \cite{Kang2001,DujovneSSC}.
\par

\begin{figure}
\includegraphics{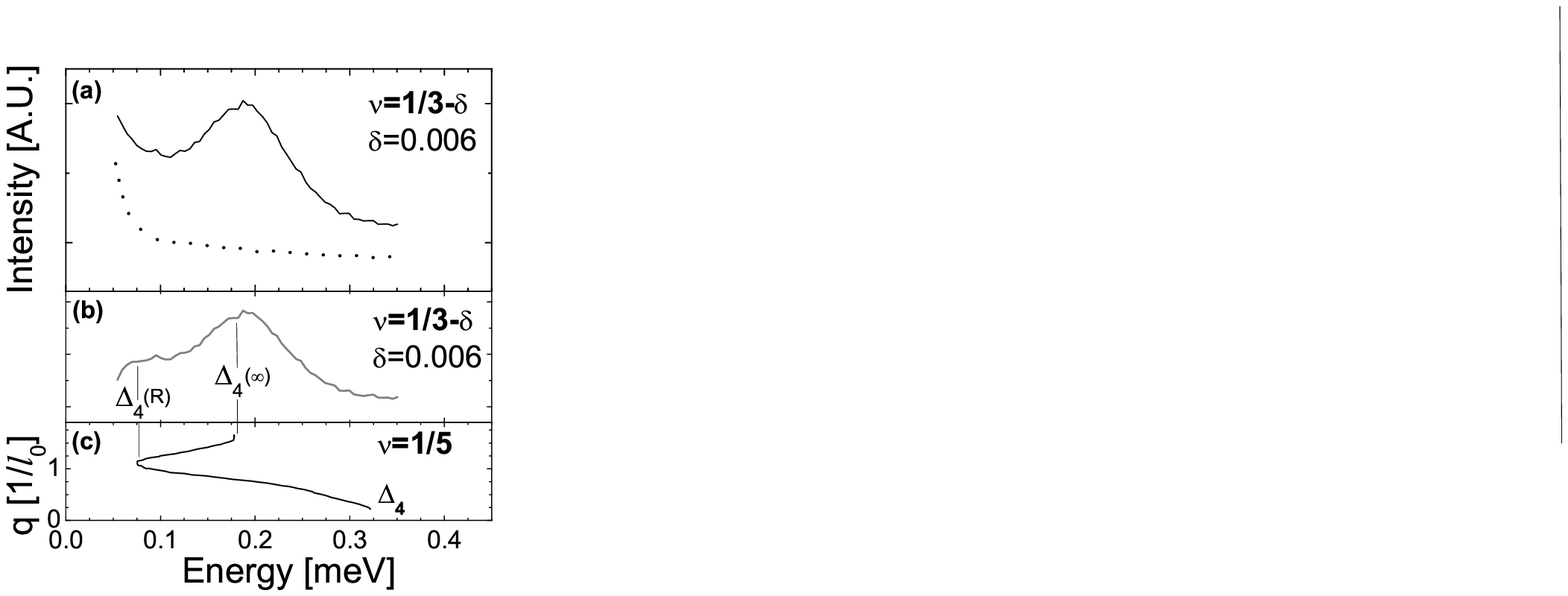}
\caption{\label{fig:figure4} (a) Spectra at $\nu=1/3-0.006$ and $T
\sim 50mK$ (solid black). Also shown is the laser tail from the
negative energy side (dotted spectrum). (b) The obtained difference
between the spectrum and the laser tail shown in part a. (c)
Dispersions of spin-conserving excitations of $^4CFs$ at $1/5$
\cite{Scarola2000b}, scaled to the density and magnetic field in (a)
and multiplied by $0.6$ to account for the effects of finite layer
width, as discussed in the text.}
\end{figure}

The low-lying excitations of the $\phi=4$ regime emerge rather
abruptly when $\nu \lesssim 1/3$ at the filling factor where
$\Delta_{2}(0)$ disappears. As seen in Fig. \ref{fig:figure4}a, the
$\Delta_{4}$ modes are already clearly seen at $\nu = 1/3 - 0.006$.
Figure \ref{fig:figure4}b shows the spectrum obtained after removal
of the overlapping tail of the laser in the spectrum of Fig.
\ref{fig:figure4}a. The corrected spectrum displays a low energy
onset at $0.07meV$ and a cut-off at higher energies. For the $^4CF$
hierarchy, the $1/3$ state is the reverse $B^*$ analog of $1/5$. The
spectrum in Fig. \ref{fig:figure4}b could be interpreted with the
calculated spin conserving $\Delta_{4}$ modes of the $1/5$ state
\cite{Kamilla&Jain,Scarola2000b}.  To account for mode softening from
the finite well width, we scale the calculated $1/5$ dispersion for
spin-conserving excitations  by a constant factor of $0.6$. This is
consistent with the correction used for $^2CF$ modes at $1/3$
\cite{Kang2001}. As shown in Fig. \ref{fig:figure4}c, the scaled
dispersion is in good agreement with the measurement. The high energy
cut-off is consistent with the large-q value of the dispersion
($\Delta_{4}(\infty)$), and the low energy onset is interpreted as
the roton of the dispersion ($\Delta_{4}(R)$). The roton may
determine the scale of the $^4CF$ temperature dependence.
\par

The scaling of $\Delta_{2}(R)$ and $\Delta_{2}(\infty)$ with $B$ seen
in Fig. \ref{fig:figure2}b goes approximately as $e^2/ \epsilon l_0$,
which is consistent with the $\Delta_{2}(0)$ and $\Delta_{2}(R)$
scaling with density \cite{Kang2001}. Better agreement is obtained
for $\Delta_{2}(\infty) = 0.95 e^2 / \epsilon l_0 (1 - L/l_0)$, where
$0.95 e^2 / \epsilon l_0$ is the value of the activation gap without
including finite width effects
\cite{Zhang&DasSarma,Nakajima&Aoki,Scarola2000a,Murthy2002}, with a
best fit of $L=37\AA=0.11w$ shown in Fig. \ref{fig:figure2}b. The
second term can be understood as the scaling of the finite width
correction with magnetic field, which in a simple approximation goes
as the ratio of a length related to the transverse extent of the 2D
electron wavefunction to $l_0$. This is consistent with calculations
of the $1/3$ activation gap that include the effects of finite width
\cite{Zhang&DasSarma,Park1998}. In simple $^2CF$ theory, the energy
level spacing at the FQH states with $\nu^*=p$ is given by
$\omega_{CF} = e |B^*| / m^*c = (C / 2p \pm 1)e^2 / \epsilon l_0$
\cite{HalperinLeeRead,Du1993}, where $C$ is a constant. Such a
scaling is consistent with that of the $\Delta_{2}(\infty)$ mode if
the condition $p=1$ is kept for all $\nu \leq 1/3$.
\par

At $\nu < 1/3$, the separation between $\Delta_{2}(R)$ and
$\Delta_{2}(\infty)$ decreases with $B$. This energy difference can
be thought of as a binding energy between the excited quasiparticle
and quasihole \cite{Kallin&Halperin1984}, arising from residual $CF$
interactions. This effect would be strongest at integer effective
$^2CF$ filling factor, where screening would be weakest. The decrease
in separation energy away from $\nu=1/3$ is interpreted as a
manifestation of increasing $^2CF$ screening of residual $CF$
interactions, although it may also include differences in the scaling
of the finite width correction at the roton \cite{Scarola2000a}.
\par

In summary, new low-lying excitations emerge in the FQH regime of
$\nu \leq 1/3$ when the quasiparticles change character from $\phi=2$
to $\phi=4$. The excitations of $^4CF$ quasiparticles are seen both
at and between the FQH effect filling factors . Coexisting with these
modes throughout the $\phi=4$ regime are spin-conserving $^2CF$
excitations seen at $\nu=1/3$. The abrupt appearance of the new
low-energy modes and the continued existence of $\Delta_{2}(R)$ and
$\Delta_{2}(\infty)$ indicate a significant change in the quantum
ground state properties. The possible coexistence of $\phi=2$ and
$\phi=4$ liquids for $\nu < 1/3$ is a topic for further study.



\begin{acknowledgments}
We wish to thank R. L. Willett for transport measurements and S. H.
Simon and P. M. Platzman for significant discussions. This work was
supported in part by the Nanoscale Science and Engineering Initiative
of the National Science Foundation under NSF Award Number CHE-0117752
and by a research grant of the W. M. Keck Foundation.
\end{acknowledgments}


\end{document}